# Interfacing a two-photon NOON state with an atomic quantum memory

Wei Zhang[1,2,†], Ming-Xin Dong[1,2,†], Dong-Sheng Ding[1,2,*], Shuai Shi[1,2], Kai-Wang[1,2], Zhi-Yuan Zhou[1,2], Guang-Can Guo[1,2] & Bao-Sen Shi[1,2,#]

[1]Key Laboratory of Quantum Information, University of Science and Technology of China, Hefei, Anhui 230026, China

[2]Synergetic Innovation Center of Quantum Information & Quantum Physics, University of Science and Technology of China, Hefei, Anhui 230026, China

[†]These authors contributed to this article equally

Corresponding author: [*]dds@ustc.edu.cn; [#]drshi@ustc.edu.cn

**Multi-photon entangled states play a crucial role in quantum information applications such as secure quantum communication, scalable computation, and high-precision quantum metrology. Quantum memory for entangled states is a key component of quantum repeaters, which are indispensable in realizing quantum communications. Storing a single photon or an entangled photon has been realized through different protocols. However, there has been no report demonstrating whether a multi-photon state can be stored in any physical system or not. Here, we report on the experimental storage of a two-photon NOON state in a cold atomic ensemble. Quantum interference measured before and after storage clearly shows that the properties of the two-photon NOON state are preserved during storage. Our experiment completes the first step towards storing a multi-photon entangled state.**

Multi-particle entangled states are the most striking states in quantum mechanics. They play a crucial role in a variety of quantum information applications from quantum communication and computation [1] to quantum metrology [1–3]. For computation, such multi-particle states provide a faithful approach to a larger-scale quantum computing. In the field of quantum metrology, the low measurement precision arising from the inevitable statistical errors in measurement can be significantly improved using multi-particle entangled states [1, 3]. Recent work reveals that a multi-particle NOON state offers advantages in high-precision phase measurements by increasing the number of particles $N$ [4–8], achieving a phase precision of $\Delta\varphi=1/N$, called the Heisenberg limit [2, 9].

The functioning of this phase sensitivity-enhanced technique using NOON-type entangled states has been demonstrated with photons [4, 6], trapped ions [10–12], and atoms [13, 14], exhibiting the significances of sub-Rayleigh resolution in quantum lithography [15], nuclear spin [5], and atomic spin waves [16]. It also points to potential applications in many fields, including microscopy [17], gravity-wave detection, measurements of material properties, and medical and biological sensing. The simplest NOON state with $N$=2 was generated experimentally in 1990 using photon pairs prepared by spontaneously parametric down conversion in a Hong–Ou–Mandel (HOM) interferometer [18–20] with a nonlinear crystal. Photonic NOON states of a large number of photons principally behave like a "Schrödinger cat", a state which has intrigued many in understanding macroscopic entanglement initially conceived by Schrödinger [21]. Because the multi-particle NOON state is interesting and essentially important for its special properties and potential applications, so its generation, transporting and storing of the multi-particle NOON state need investigating.

Quantum memory requires storing and retrieving photonic quantum states on demand. Storing a single photon or an entangled photon has been realized by many groups [22–30] through different storage protocols. Because storage efficiency is low, and control of the purity and strict requirements in bandwidth matching between multi-photons and memory are difficult, there has been no progress in the quantum storage of the photonic NOON states in any physical system to the best of our knowledge. Here, we report on the experimental storage of a two-photon NOON state using the Raman storage protocol [31]. The state is generated via the spontaneous Raman scattering (SRS) process in a cold atomic ensemble using a HOM interferometer. The measured interferences clearly show that the properties of the two-photon NOON states are conserved in the storage, providing evidence that the photonic NOON state is actually stored in our atomic ensemble memory. All results infer the capability of storing multi-photon entangled states for quantum memory that is essential for quantum information processing.

**Results**

**Generation of photonic NOON states.** We first prepare non-classical photon pairs via the SRS process in an optically thick atomic ensemble of Rubidium ($^{85}$Rb) trapped in a two-dimensional magneto-optical trap (MOT) [32]. In the experimental set-up (figure 1), a double-Λ energy level configuration, which consists of three states |1>, |2>, and |3>, corresponding to atomic levels $5S_{1/2}(F=2)$, $5S_{1/2}(F=3)$, and $5P_{1/2}(F=3)$, is used to generate photon pairs. The photon of Signal 1, generated with the writing of Pump 1, is collected into a 50-m long fibre with a ~80% coupling efficiency. After elapsed time $\Delta T$, Signal 2 was read out from the atomic ensemble by applying the beam of Pump 2, and was collected into a 2-m long fibre. Both Signal 1 and Signal 2 photons are then inputted into a beam splitter (BS1) from different ports. With a suitable selected detuning of Pumps 1 and 2, a best-fit $\Delta T$, and the polarization adjustment (horizontal polarization), the two photons are completely indistinguishable at BS1. Under these circumstances, a two-photon NOON state is generated at outputs of BS1. We next store this NOON state in MOT B using an active-locked Mach–Zehnder interferometer. We recorded the interference patterns from the outputs of BS2 and BS3 before and after storage with a phase change caused by a phase module.

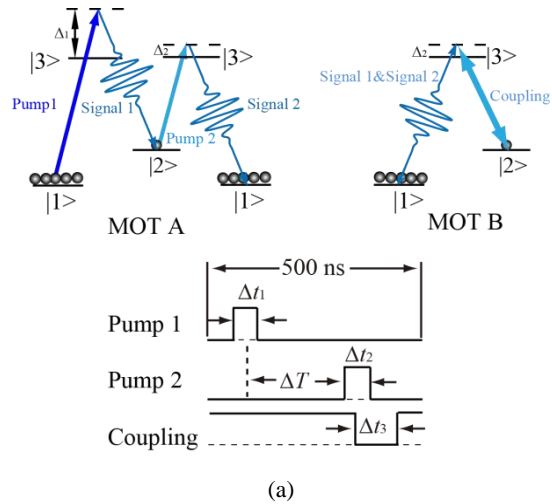

(a)

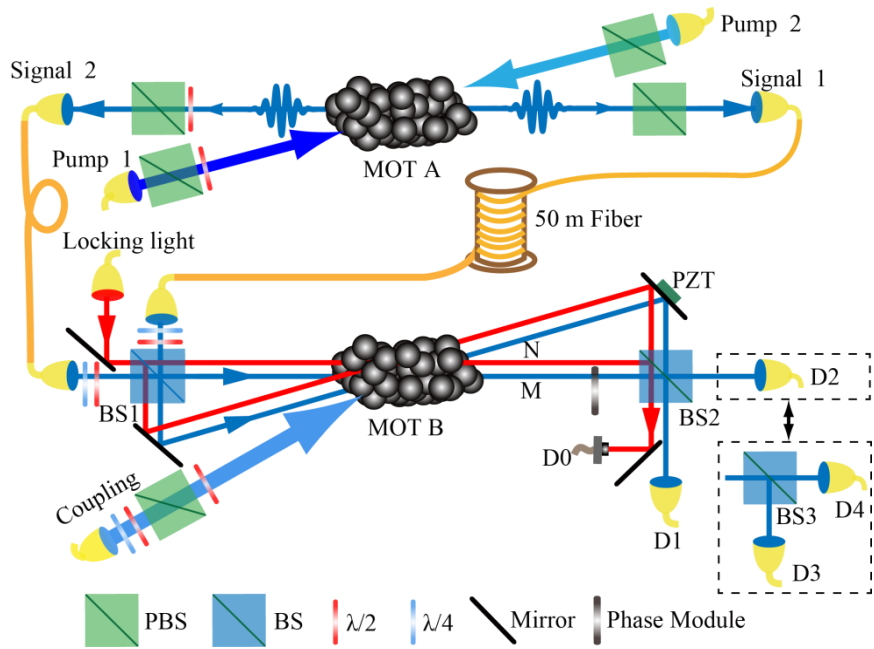

(b)

**Figure 1 | Generation and storage of a two-photon NOON state.** (a) Energy diagram and timing sequence. Both Pumps 1 and 2 are pulses with durations of $\Delta t_1 = \Delta t_2 = 70$ ns. The time delay of Pump 2 is $\Delta T$, and the storage time in MOT B is set by $\Delta t_3$. $\Delta_1$ and $\Delta_2$ represent single-photon detuning of the Pump-1 and Pump-2/Coupling light. Here $\Delta_2 = 70$ MHz, and $\Delta_1 - \Delta_2$ equals the energy difference between |1> and |2>. (b) Simplified experimental set-up. PBS: polarizing beam splitter; BS: beam splitter; λ/2: half-wave plate; λ/4: quarter-wave plate; D0: photodetector; D1 (D2/D3/D4): single-photon detector (Avalanche diode, PerkinElmer SPCM-AQR-16-FC); PZT: piezoelectric ceramics; *M/N*: represents the memory path. Pump 1 counter-propagates with Pump 2 through the atomic ensemble in MOT A. Photons of Signal 1 and Signal 2 are collected with 1.5 ° to the direction of pumping light. The coupling light is obliquely incident on the atomic ensemble in MOT B with the same angle of 1.5 ° with respect to memory paths *M* and *N*. The power for the light beams (Pump 1, Pump 2, and Coupling) is about 100 mW, 20 mW, and 24 mW, respectively.

Our system works periodically with a repetition rate of 100 Hz including an 8.6-ms trapping and initial state preparation time, and a 1.4-ms operation time comprising 2800 cycles with a cycle time of 500 ns. In each cycle, Pumps 1 and 2 are pulsed by acousto-optic modulators. The waist for two pump laser beams in the centre of atomic ensemble is ~2 mm. The photons of both Signals 1 and 2 are collected/focused through a lens [not depicted in figure 1(b)] with a focal length of *f*=500 mm in MOT A/B. The optical depth of the atomic ensemble in MOTs A and B is about 20 and 50, respectively.

To generate a two-photon NOON state, we firstly performed HOM interferometry after BS1 using detectors D1 and D2 with BS2 removed and MOT B blocked. A typical HOM-dip is observed [figure 2(a)]. After that, we moved BS2 back and blocked path *N* (MOT B remaining blocked). In contrast, a peak is observed after BS2 [figure 2(b)].

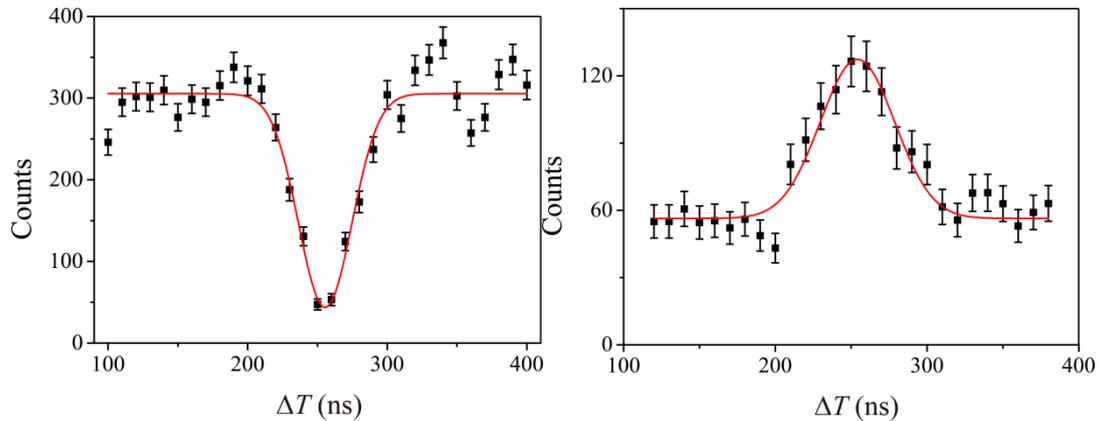

(a)             (b)

**Figure 2 | Observation of interference dip and interference peak in Signals 1 and 2.** Coincidence counts rates of D1 and D2 taken over 200 s (a) without and (b) with BS2 with increasing $\Delta T$ (time step:10 ns). The black squares are experimental data; red curve is a theoretical fit using a Gaussian function. Error bars represent ±standard deviation (s.d.) estimated from Poisson statistics.

The visibility of HOM interference is ~83% with a defined visibility formula of $V_0=(C_{max}-C_{min})/C_{max}$, where $C_{max}$ and $C_{min}$ correspond to the maximum and minimum counts rate, respectively. The peak in figure 2(b) indicates that the Fock state $|n=2\rangle$ is generated in path $M$, as the experimental data are consistent with the ideal situation in which $g^2_{max}=2$. At this point, the photonic NOON state is indeed generated; it is denoted by

$$|\psi_0\rangle=(|2\rangle|0\rangle+|0\rangle|2\rangle)/\sqrt{2}.$$

**Quantum storage of photonic NOON states.** Next, we focus on the key part in our experiment—quantum storage of the two-photon NOON state in our atomic-ensemble memory. The two-photon NOON state generated, consisting of two-photon states along paths $M$ and $N$, is focused on the centre of the atomic ensemble in MOT B for storage. The stored atomic NOON state is denoted as

$$|\psi_1\rangle=(|2_M\rangle|0\rangle+|0\rangle|2_N\rangle)/\sqrt{2},$$

where $|2_M\rangle$ and $|2_N\rangle$ represent the two spin waves simultaneously existing in paths $M$ and $N$, respectively

The photonic NOON state is verified using an active-locked Mach–Zehnder interferometer embedded with the atomic ensemble in MOT B and a phase module with phase angle recorded as $\alpha$ [see Methods]. A theoretical calculation indicated that the interference phenomenon would be observed at the output of the Mach–Zehnder interferometer with the change of phase angle $\alpha$ when the NOON state is input. This is a standard method in determining the number of particles in the NOON state. For a two-photon NOON state, the output pattern from the interferometer, which is measured by the coincidence counts of detectors D1 and D2 can be expressed as [33],

$$R_{NOON} \propto (1-cos(2\alpha))/2.$$

Here we measure the coincidence rate between D1 and D2 with the change in phase angle $\alpha=2\theta$

for the NOON state before and after storage [see Method]. To highlight NOON-state properties, we further checked the single-photon quantum interference for comparison. In measuring this interference, Signal 2 is directly detected as a trigger without being directed to BS1, and Signal 1 is directed through BS1 to MOT B for storage and detection by D2. The ideal interference pattern for single photons has the form of

$$R_{single} \propto (1-cos(\alpha))/2,$$

According to this expression and condition $\alpha=2\theta$, the interference for a two-photon NOON state exhibits a $R_{NOON} \propto (1-cos(4\theta))/2$ dependence; for a single photon, the dependence is $R_{single} \propto (1-cos(2\theta))/2$. Figure 3 shows the measured interference curves for single-photon states and NOON states before and after storage in MOT B. We find the experimental data are consistent with the calculated theory.

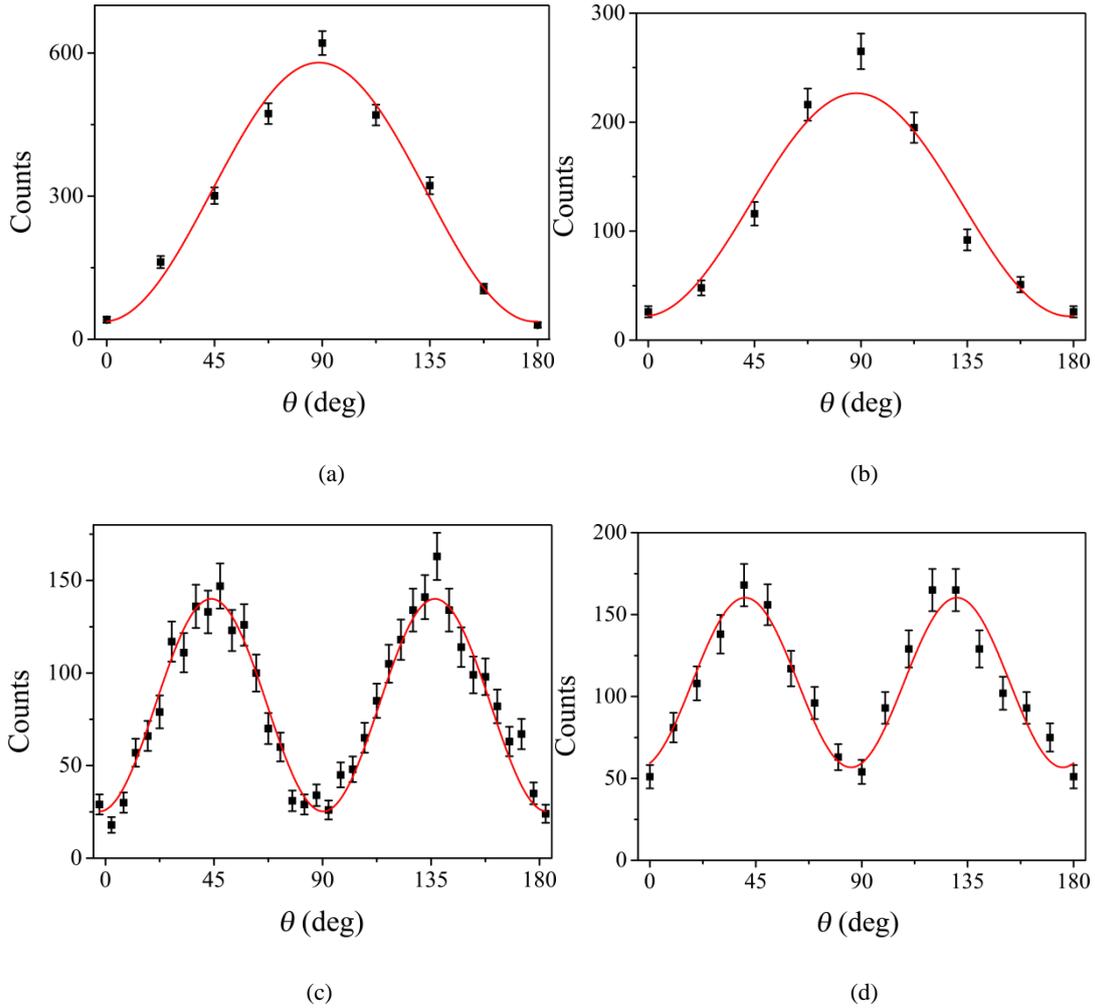

(a)

(b)

(c)

(d)

**Figure 3 | Interference curves at BS2.** Interference curves for single-photon states (a, b) and two-photon NOON

states (c, d) before (a, c) and after (b, d) storage at BS2. Error bars represent ±s.d estimated from Poisson statistics. Red curve is a fit using Cosine function.

The interference for single photons shows visibilities of 90.9±1.7% before storage [figure 3(a)] and 82.0±3.2% after storage [figure 3(b)]; the interference of the two-photon NOON states show visibilities of 74.2 ±3.8% before storage [figure 3(c)] and 54.2±4.0% after storage [figure 3(d)]. At this stage, the visibility is defined as $V_1=(C_{max}-C_{min})/(C_{max}+C_{min})$.

Here the low visibility of the interference for the two-photon NOON state after storage results from the low storage efficiency of the state. Assuming the efficiency for storing a single photon is $\beta$, the storage efficiency for a two-photon NOON state is $\sim \beta^2$. Here $\beta \sim 20\%$ [31], and hence the measured storage efficiency is ~4% for two-photon NOON state. Under these conditions, random coincidences coming from experimental noise and dark counts lower the visibility of interference patterns.

Indeed, at this moment, we are not able to determine the input state according to the interference curves coming from coincidences in D1 and D2. As shown in ref. 33, the shape of the interference curves for two-photon NOON states and coherent states (input from one port of BS1) are the same. To further distinguish the input state, we next replaced the D2 with BS3, D3, and D4 [see figure 1(b)]. From ref. 33, the interference curves coming from D3 and D4 are distinct for the two-photon NOON states and for the coherent states (input from one port of BS1). For coherent states, the pattern of interference becomes $R'_{coherent} \propto (1-cos(2\theta))^2/4$ whereas for two-photon NOON states it is still $R'_{NOON} \propto (1-cos(4\theta))/2$. Thus, we can identify the input state as the two-photon NOON state according to the interference curve coming from the coincidence of D3 and D4 (figure 4).

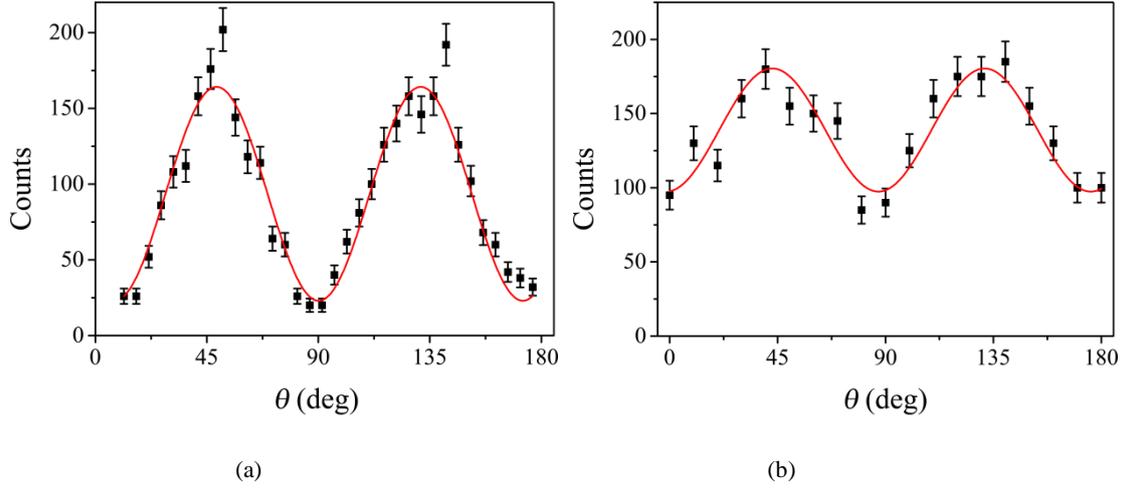

**Figure 4 | Interference curves at BS3.** (a) and (b) are interference curves for the two-photon NOON state before and after storage. Error bars represent ±s.d. estimated from Poisson statistics. Red curve is a fit using Cosine function.

From figure 4, with the initial phase redefined as being the same, we can see that the interference curves have the same shape with those of figure 3(c) and (d), confirming two-photon NOON state's generation and storage.

**Discussion**

We experimentally generated a two-photon NOON state based on an atomic ensemble, and successfully achieved quantum storage of this NOON state in another atomic ensemble. We found that this NOON state has no collective effect when stored in the atomic ensemble, which means that photons are stored individually and independently as far as ascertained with the storage efficiency obtained. This may be a challenge for storing multi-particle entanglement with large number of particles. However, it would not be a problem with the increasing of storage efficiency in the future.

In 2012, holographic storage of entangled biphotons was achieved in which a BELL-type entanglement was stored [35]. Our experiment achieves the first multi-photon NOON-type storage using atomic ensembles. Recently, the generation of multi-particle states was explored in a wide variety of systems, and by the end of 2016, ten-photon entanglement was experimentally generated [34]. Rapid developments for applications in this field demand the effective control of this state. In this regard, the quantum storage of multi-particle state is very important for future practical devices.

## Methods

### Phase module

A phase module, consisting of two quarter-wave plates interposed by one half-wave plate, is used to change the relative phase between paths *M* and *N*. The transfer matrices of the quarter-wave plate and half-wave plate are

$$U_{QWP} = \begin{pmatrix} i-\cos(2q) & \sin(2q) \\ \sin(2q) & i+\cos(2q) \end{pmatrix}/\sqrt{2},$$

$$U_{HWP} = \begin{pmatrix} \cos(2\theta) & -\sin(2\theta) \\ -\sin(2\theta) & -\cos(2\theta) \end{pmatrix},$$

for which $q$ and $\theta$ are the angles of the fast axes with respect to the vertical. With the arrangement $q_1=\pi/4$ and $q_2=\pi/4$ for the first and second quarter-wave plates and $\theta$ for the middle half-wave plate, light $E_0|H\rangle$ changes to $-e^{i2\theta}E_0|H\rangle$ after passing through the phase module; here $|H\rangle$ represents the horizontally polarized light. Hence the phase angle produced by this module is $\alpha=2\theta$.

### Detecting two-photon NOON state after storage

As mentioned in the main text, detectors D1 and D2 record the interference counts of the NOON states before and after storage. For interference before storage, this is satisfactory but for interference after storage, coincidences in D1 and D2 counts cannot distinguish between a leaked signal and a stored signal. We further applied a synchronized GATE signal that filtered out only the stored signal for coincidence measurements.

## Acknowledgements

This work was supported by the National Natural Science Foundation of China (Grant Nos. 61275115, 61435011, 61525504, and 11604322).

## Author contributions

D.-S.D. conceived the experiment. The experimental work was conducted by M.-X.D. and W. Z.,

data analysis is done by W. Z. with assistance from S.S., K.W., S.-L.L. , and Z.-Y.Z. Moreover, W.Z. and D.-S.D. wrote this paper with assistance from B.-S.S., and B.-S.S. and G.-C.G supervised the project.

**Competing financial interests**

The authors declare no competing financial interests.